# Strong Spreading in a Droplet Flow for Low-Dimensional Nanostructures Growth

Zheng Fan[1,3,4], Jean-Luc Maurice[1], Ileana Florea[1], Wanghua Chen[1,5], Linwei Yu[1,6], Stéphane Guilet[2], Edmond Cambril[2], Xavier Lafosse[2], Laurent Couraud[2], Kamel Merghem[2], Sophie Bouchoule[2], Pere Roca i Cabarrocas[1]

[1]Laboratoire de Physique des Interfaces et des Couches Minces, CNRS, École Polytechnique, Institut Polytechnique de Paris, 91128 Palaiseau, France
[2]Centre de Nanosciences et de Nanotechnologies, CNRS, Université Paris-Sud, Université Paris-Saclay, Avenue de la Vauve, 91120 Palaiseau, France
[3]Tianjin International Center for Nanoparticles and Nanosystems, Tianjin University, 300072 Tianjin, P. R. China
[4]ShanghaiTech Quantum Device Lab, ShanghaiTech University, 201210 Shanghai, P. R. China
[5]School of Physical Science and Technology, Ningbo University, Ningbo 315211, P. R. China
[6]School of Electronic Science and Engineering, Nanjing University, 210096 Nanjing, P. R. China

**Abstract**: We report an *in situ* transmission electron microscopy observation of an indium droplet flowing on a silicon nitride membrane with a coating layer of hydrogenated amorphous silicon (*a*-Si:H), with the production of in-plane *c*-Si nanowire in its trail. We observe that the droplet strongly spreads on the *a*-Si:H coated surface while it dewets from the *c*-Si NW. This *in situ* observation, combined with the geometric analysis of such liquid-solid systems, presents nice consistency with de Gennes' theoretic prediction of the droplet hydrodynamics steered by reactive spreading, where the wettability gradient for the droplet flowing is maintained by a progressively autophobic process due to the droplet mediated crystallization of *a*-Si:H. Interestingly, we record temperature dependent evolution of the droplet-nanowire interface, which leads the droplet break-up, self-turning and the nanoflake-to-nanowire transition. We elucidate these rich nanofluidic phenomena by a model based on the heterogeneous nucleation governed reactive spreading.

Droplet based micro/nanofluidics has been demonstrated as a versatile tool in many applications. [1-9] Metal droplets are often applied to mediate nucleation and crystal growth, which dates back to Wagner's pioneering work on the vapor-liquid-solid (VLS) growth of Si whiskers by gold. [10] In particular, direct writing of in-plane low-dimensional nanostructures arises since the last decade, which relies on crawling metal droplets as catalytic media. [11] Substantial efforts have been made on the controllable manipulation of catalytic droplets on patterned substrates like nanofacets [12-14], guiding steps [15-17], or nanochannels [18]. Recently, VLS-grown in-plane $MoS_2$ nanoribbons have been realized by the moving Na-Mo-O droplets, which avoids the dry etching process to pattern the 2D sheets. [19] In a reversal manner, Ni droplets can migrate on graphene sheets in $H_2$ atmosphere and tailor them into nanoribbons having a width below 10 nm. [20]

In general, a wettability gradient on a solid surface is required for the droplet self-propulsion, where unbalanced Young's force drives the droplet moving towards a more wettable area. [21] Various techniques have been reported to activate and maintain the droplets' surface migration, where videos were recorded as direct evidences. [22-26] When the droplets shrink down to nanometer scale, *in situ* transmission electron microscopy (TEM) exhibits its unique advantages, as it can directly reveal the nucleation events, as well as the dynamics at interfaces between droplets and nanocrystals. [27-32] However, visualization is still in high demand for understanding the growth of in-plane low-dimensional nanostructures mediated by the flowing droplets, since their nanofluidic behavior predominantly determines the growth of the nanostructures. [17,32,33]

To this aim, we selected the flowing indium (In) droplet for the production of in-plane solid-liquid-solid (IPSLS) *c*-Si nanowire (NW) as a platform [33,34], to explore the dynamics of the droplet transport by *in situ* TEM technique (The experiments are detailed in the Supplementary Materials (SM) Section 1). The *in situ* observation started from a *c*-Si NW grown on a silicon nitride ($SiN_x$) membrane in the TEM environment, with a solid In NP at the NW end (see Fig. S7). Once the membrane was heated to 350 °C, a rich interplay between the In droplet, the *a*-Si:H coated membrane and the *c*-Si NW arose. Fig. 1(a) displays a sequence of TEM images extracted from Video SV1. The In droplet at the end of the as-grown *c*-Si NW (see the blue part) started to melt and dissolve the *a*-Si:H layer at *t*=2 s. The thin In thin film along the front edge of the droplet, in contrast with the thick droplet bulk, first advanced laterally into two wings from *t*=4 to 6 s, which were pinned by the two arms (marked by the two yellow dashed circles, probably the *c*-Si precipitates during the In droplet solidification, see Fig. S5). The droplet slipped forward with the liquid-solid (L-S) contact drastically stretched, during which new phases of *c*-Si kept being precipitated (see the red part). However, the droplet managed to escape from the arms with its wings rapidly relaxing from *t*=10 to 14 s, meanwhile dissolving the *a*-Si:H along its relaxation path (see the red arrows), uncovering the footprint of the droplet trail whose borders are marked by the black dashed lines (we name it trench as seen in Fig. 3(a)), and producing a new segment of NW with three facets in contact with the droplet (see the yellow part). As no thin wetting layers were observed at the droplet-NW interface, this Si phase transition induced droplet dewetting is consistent with de Gennes' theoretical prediction of droplet hydrodynamics driven by reactive spreading. [35] In other words, the $SiN_x$ membrane recovered its *indium-phobicity* after the *a*-Si:H was swept by the droplet.

Moreover, the instabilities of the moving droplet were also recorded. The droplet-NW interface gradually developed three facets at *t*=14 s. By superposing the blue, red and yellow parts of the NW at *t*=14 s in the following TEM images, one can see that *c*-Si mainly precipitated normal to the facet 2 at *t*=15 s, which brought about a risk of squeezing the droplet into two. Afterward, the facets 1 and 2 moved forward until *t*=19 s, causing a remarkable elongation of the droplet-NW contact line that finally transferred to the irregular morphology of a faceted NW. Hereafter, the facet 3 started to advance and the right wingtip of the droplet was pinched off so that the droplet relaxed and accumulated on the left side (*t*=20 to 22 s). This phenomenon occurred a second time later, where the faceted droplet-NW interface (see *t*=50 s) finally led to a growth of a *c*-Si nanoflake (denoted as NF, red part) rather than NW (blue part), as seen in Fig. 1(b). Interestingly, the droplet broke up twice during its motion, at *t*= 21 s and 61 s. Such a phenomenon also occurs in an IPSLS process with tin (Sn) as catalyst, during which the Sn droplet keeps losing its mass which consequently results in the formation of tail-like Si or Ge NWs. [36,37] More detailed discussion on the zigzag growth of Sn catalyzed IPSLS Si or Ge NWs is provided in the SM Section 4. This droplet break-up behavior is a ubiquitous signature during the deformation of viscous droplet flow, which could be attributed to the competition between the shearing force within the droplet and the droplet surface tension. [38,39]

Surprisingly, when lowering the temperature to 300 °C, the droplet steadily sat on the membrane, with a wetting layer along the L-S contact line while no *c*-Si precipitated (see Fig. 2(a), *t*=0 s archived in Video SV2). No significant In wetting behavior was observed in the trench (contoured by the black dashed line) nor on the NW. After ramping the temperature to 400 °C, the droplet spread into a fan-like shape and its migration kicked off (see Fig. 2(b)). In contrast with the observation at 350 °C, the droplet's shape and the droplet-NW interface were fairly maintained, so the variation of NW diameter was eliminated to a large extent. However, the droplet's moving direction changed frequently (see *t*=19 s to 23 s). This may be associated with the asymmetric wetting profile which could be attributed to the chemical inhomogeneity of the *a*-Si:H layer, where the droplet spread to the more wettable surface. [23] Another interesting case of droplet turning (see Fig. 2(c)) occurred once the In droplet was trapped in a situation where the *a*-Si:H was only present to its right side (*t* =45 s), it spread and turned towards the *a*-Si:H covered surface (*t*=46, 47 s). This self-avoiding of its moving trace is a significant feature of the droplet motion maintained by progressive reactive spreading. [35]

One can see that a NF is easier to grow at 350 °C by a slow and drastically deformed droplet (～20 nm $s^{-1}$) in comparison with that in a fast (～100 nm $s^{-1}$) stick-slip motion at 400 °C. Such a flake-to-fiber transition ubiquitously exists in the solidification of Al-Si eutectic alloys, where fast growth favors the formation of Si fibers. [40] Apparently, the lower temperature retards (at 350 °C, see Fig. 1) or even suppresses (at 300 °C, see Fig. 2(a)) the Si diffusion into the In droplet and the following Si nucleation, which slows down the droplet motion. However, the temperature dependent droplet reactive spreading needs to be elucidated, which predominantly affects the morphologies of the Si precipitates. Since the droplet surface tension varies little in such a narrow temperature window [41] (the calculation is seen in SM Section 2), the chemical reaction between In and *a*-Si:H should be accounted. In order to activate the Si precipitation from In droplet, the Si chemical potential difference between the supersaturated and saturated In droplet $\Delta\mu$ should overcome the heterogeneous nucleation barrier $\Delta G^{*}_{hetero}$:

$$\Delta\mu_{Si} > \Delta G^{*}_{hetero}. \tag{1}$$

Being a process of heterogeneous nucleation and crystal growth from solution [42], Equation (1) can be transformed into

$$C_{s,Si(In)} > C_{0,Si(In)}\exp\left[\sqrt[3]{\chi f(\theta_{Si-Sub})}\Omega_{Si}\gamma_{Si-In}/(k_B T)\right], \tag{2}$$

where $C_{s,Si(In)}$ and $C_{0,Si(In)}$ denote the Si concentration in the In droplet in a supersaturated state and in the equilibrium state, respectively, and $C_{0,Si(In)}$ is considered constant from 350 °C to 400 °C [43]; $\chi$ the shape factor of the Si nucleus, $\Omega_{Si}$ the volume of a Si atom, $\gamma_{Si-In}$ the interfacial energy between the Si nucleus and the liquid In, $k_B$ the Boltzmann constant, $T$ the growth temperature, $f(\theta_{Si-Sub})$ the contact angle ($\theta_{Si-Sub}$) factor between the Si nucleus and the substrate, which increases monotonically with the increment of $\theta_{Si-Sub}$ [42]. A more detailed deduction is provided in the SM Section 3. Thereby, when the growth temperature $T$ decreases, the evolution of the droplet reactive spreading tends to develop in two aspects: (1) on the one hand, before each step forward, the In droplet has to spread more aggressively to cover more *a*-Si:H area and feed itself more Si atoms, so that $C_{s,Si(In)}$ will be raised and a higher Si supersaturation degree (i.e. $C_{s,Si(In)}/C_{0,Si(In)}$) will be approached to overcome the nucleation barrier $\Delta G^*_{hetero}$, which consequently renders the formation of the multi-faceted droplet-NW interfaces; (2) on the other hand, the Si nucleus tends to wet the substrate (i.e. $\theta_{Si-Sub}$ decreases) so that $f(\theta_{Si-Sub})$ will drop to compensate the lowered $T$, which gives rise to the growth of flattened *c*-Si (i.e. a NF) rather than a NW (see Fig. 3(b)).

As a complement to the *in situ* TEM observations, we also carried out a geometric analysis of IPSLS *c*-Si NWs grown on $SiO_2$ substrates (see in SM Section 1), which provides additional information of the In droplet reactive spreading behavior. Fig. 3(a) shows a scanning electron microscope (SEM) image of a typical IPSLS *c*-Si NW grown on $SiO_2$, with a solidified In nanoparticle (NP) and a trench that records the trace of In droplet migration. Direct evidence on the trench structure was obtained by performing high-angle-annular-dark-field scanning transmission electron microscopy (HAADF-STEM) energy-dispersive X-ray spectroscopy (EDS) chemical map on a cross-section of the NW prepared by focused ion beam (FIB), as seen in Fig. 3(b). The boundaries of the *a*-Si:H layer are marked by a white dashed line, where $h_{aSi}$ denotes the thickness. One can see that the NW is located at the trench center with *a*-Si:H completely exhausted, while a small amount of residual *a*-Si:H remains at the edges of the trench. A top-view scheme of a NW is illustrated in Fig. 3(c), labeled with geometric parameters $d_{NP}$ for the solid In NP diameter, and $d_{In\text{-}aSi}$ for the trench width which can be considered as the L-S contact width. A linear fitting for exploiting the relationship between $d_{NP}$, $d_{In\text{-}aSi}$ and $h_{aSi}$ is provided in Fig. 3(d):

$$W = 1.26 + 3.2H, \quad (3)$$

where $H$ denotes $h_{a\mathrm{Si}}/d_{NP}$ (i.e. the relative thickness of *a*-Si:H to the droplet size), and $W$ denotes $d_{In\text{-}aSi}/d_{NP}$ (i.e. the relative trench width to the droplet size). The intercept of 1.26 indicates that if there is no *a*-Si:H coating layer (i.e. $H$ approaches 0), $W$ equals 1.26 (i.e. $d_{In\text{-}SiO2}=1.26d_{s\text{-}NP}$). The deduction of this relationship is seen in SM Section 5. Considering a constant mass of a spherical NP before and after melting, this ratio implies that an In droplet on a bare $SiO_2$ surface is a semi-sphere, with a contact angle $\theta_{SiO2}$ of 90° [44]. However, once the $SiO_2$ surface is coated by *a*-Si:H, the L-S contact line is stretched in width (i.e. $d_{In\text{-}aSi} > d_{In\text{-}SiO2}$), therefore the contact angle shall be lowered (i.e. $\theta_{aSi} < 90°$), which fairly demonstrates that *a*-Si:H coating layer is *indium-philic*. Moreover, since $W$ increases monotonically with increasing $H$, Equation (3) indicates that a relatively thicker *a*-Si:H renders stronger spreading of the droplet (i.e. the relative trench width *W* is broader on relatively thicker *a*-Si:H coated substrate). The dependence of the *a*-Si:H thickness on the droplet reactive spreading is discussed in the SM Section 5. Fig. 4(a) illustrates the evolution of an In NP from solid to liquid, and from non-reactive spreading on bare $SiO_2$ to a reactive mode once an *a*-Si:H layer is coated on the $SiO_2$ surface.

Combining our *in situ* TEM observations with the geometric analysis, we are able to draw a general picture of a flowing In droplet for an IPSLS *c*-Si NW growth. As depicted in Fig. 4(b), a continuous NW growth is mainly enabled by three cyclic events: (i) the In droplet reactively spreads on the *a*-Si:H coated substrate surface with higher wettability ($\gamma^+$) where liquid In alloys with Si atoms dissolved from *a*-Si:H until supersaturation of Si within the droplet, which thereafter activates heterogeneous nucleation of *c*-Si and thus lowers the substrate surface wettability ($\gamma^-$). Such process is driven by the Gibbs free energy difference between *a*-Si:H and *c*-Si [45]; (ii) consequently, a wettability gradient is established, from the non-reactive bare substrate surface (i.e. $SiO_2$, $Si_3N_4$, etc.) with precipitated *c*-Si to that coated by reactive *a*-Si:H; (iii) the In droplet slips forward to the high $\gamma^+$ surface, which thereby maintains the droplet contact with *a*-Si:H. However, a *c*-Si NW is not the only type of product by a flowing droplet. As seen in Fig. S4, for large size drops in micrometer scale, as there will be more sites for the Si precipitation in such cases, separated Si crystals will be produced rather than a NW. In addition, it has to be mentioned that our *in situ* observation of droplet reactive spreading also brings deep insights into the VLS growth process in many aspects, which affects the NW tapering [46-48] and determine the VLS growth of lateral NWs [18,49] (see Figs. S8-S12).

To summarize, we have performed an *in situ* TEM study of in-plane *c*-Si NW growth led by a self-propelling In droplet. Geometric analysis unveils a stronger droplet spreading behavior on *a*-Si:H coated substrate than on a bare substrate ($SiN_x$ membrane or $SiO_2$). *In situ* TEM observations, reveal drastic deformation of the In droplet with the wetting layer at its advancing edge, which demonstrates a reactive spreading behavior. The In droplet flowing was powered by a progressively autophobic process, that is, the droplet dewets the substrate surface once the *a*-Si:H is dissolved and transformed into *c*-Si, which enables a continuous NW growth along its trail. Random turning of the flowing droplet led to a zigzag growth of NWs, probably due to the heterogeneity of *a*-Si:H. Moreover, we observed a growth of faceted *c*-Si NF rather than NW when the droplet crept slowly (20 nm $s^{-1}$), whose morphology was transferred from the elongated and faceted droplet-NW interface. The suppression of this irregular growth was enabled at a high growth rate (100 nm $s^{-1}$) by increasing the temperature, since higher temperature enables the Si nucleation from In droplet in a lower supersaturation degree, which does not demand drastic droplet reactive spreading to accommodate more Si atoms.

This work was partly supported by the French RENATECH network and the French National Research Agency (ANR) through the TEMPOS Equipex project (ANR-10-EQPX-0050), poles NanoMAX and NanoTEM. Z. F. acknowledges the Chinese Scholarship Council and FX-conseil for funding his PhD.

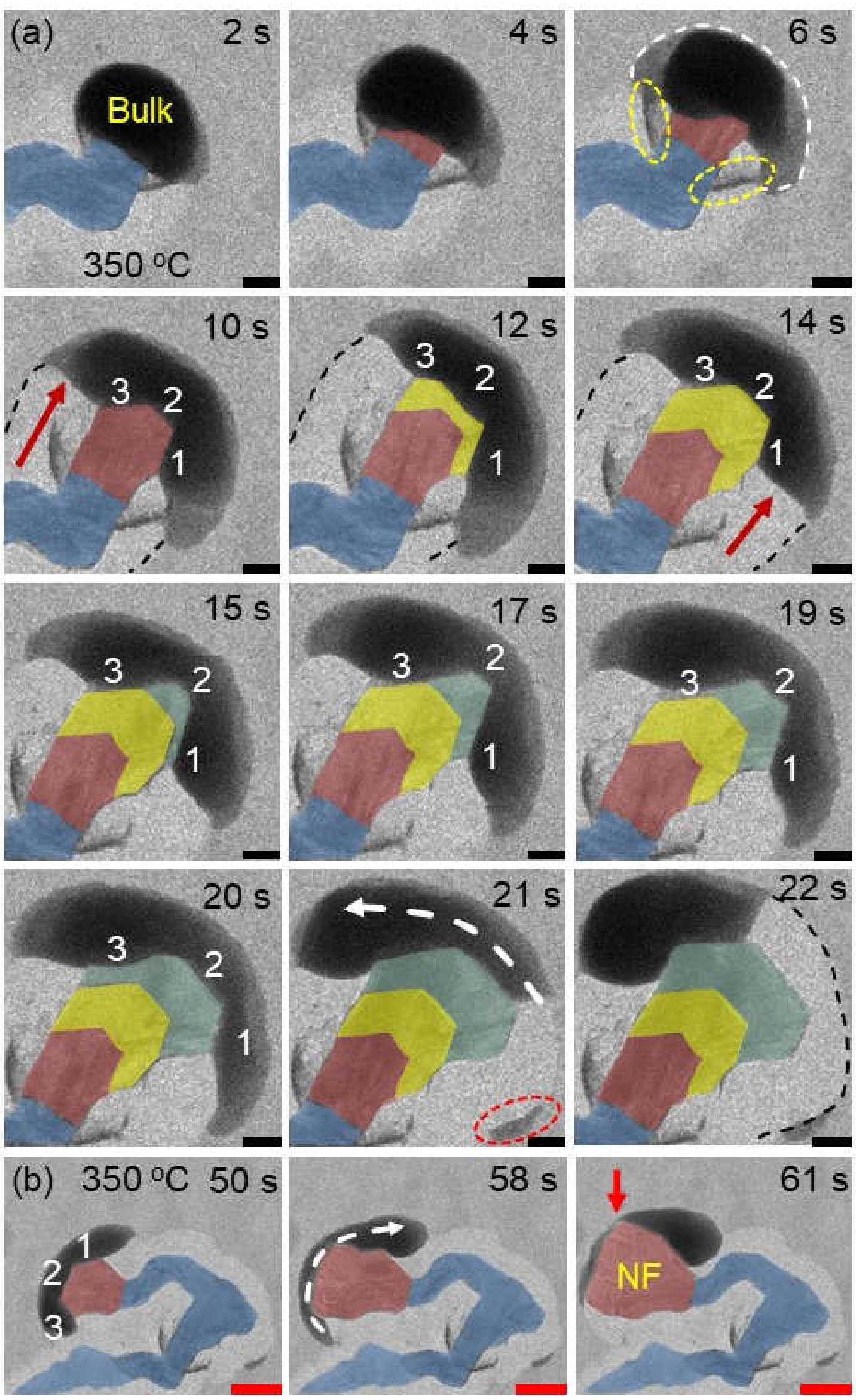

FIG. 1. Two sequences of TEM images extracted from Video S1 of the IPSLS *c*-Si NW growth at 350 °C: (a) the initial stage of reactive spreading of *a*-Si:H coated $SiN_x$ membrane by a deformed In droplet (from *t*=2 to 14 s), and the growth of NW in irregular morphology due to the faceted droplet-NW interface with the droplet break up (from *t*=15 to 22 s); (b) a growth of *c*-Si NF rather than NW that results from the extensive stretching of In droplet along the faceted droplet-NW interface. The scale bars are 100 nm, 400 nm in (a) and (b), respectively.

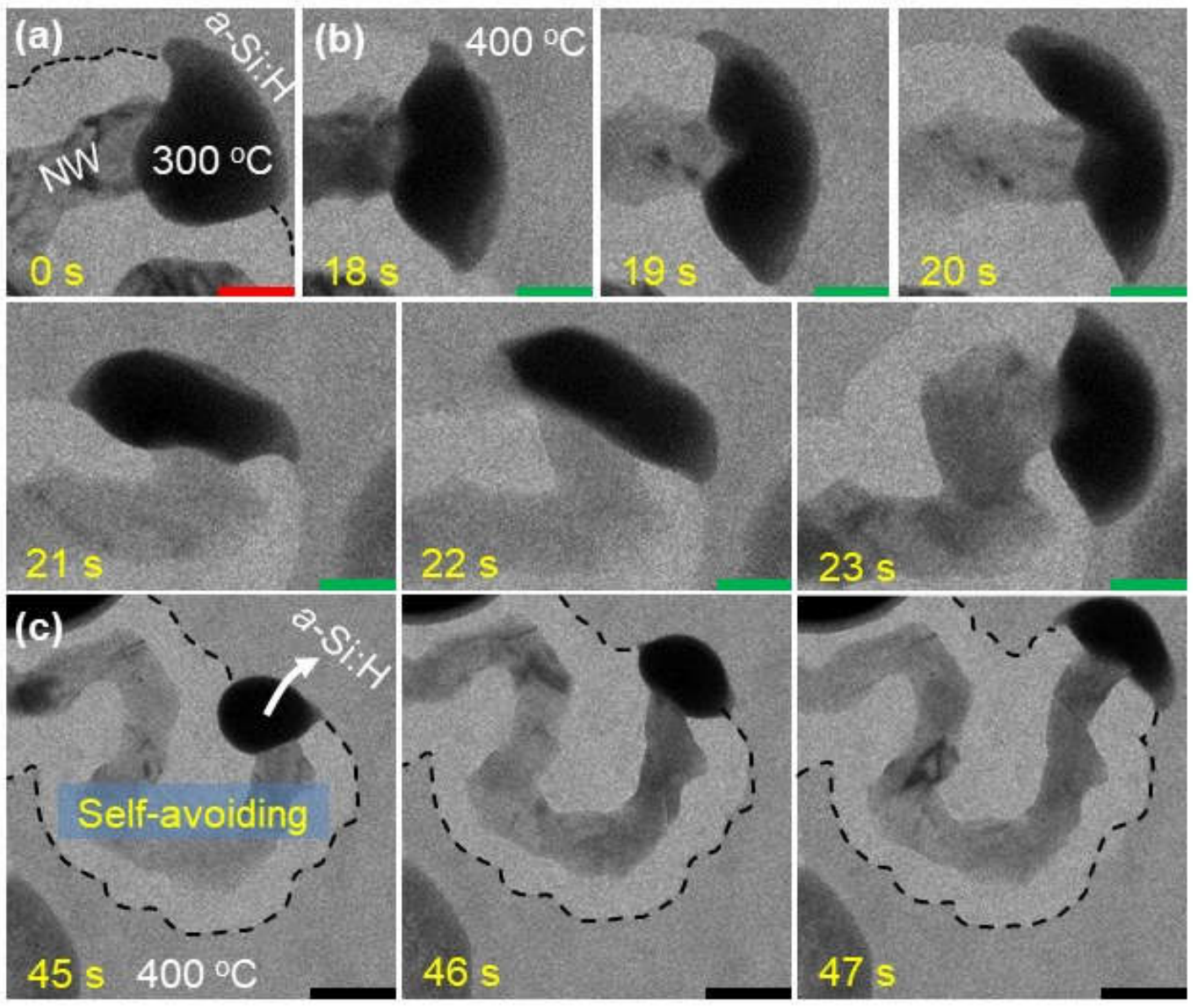

FIG. 2. TEM images extracted from Video S2. (a) In droplet steadily sat on the membrane at 300°C, with a thin wetting layer on *a*-Si:H. (b) When the temperature reached 400°C, the droplet reactively wet the *a*-Si:H coated membrane surface and started to move randomly with the production of a NW with a relatively uniform diameter. (c) Droplet self-avoiding its trace: once touching the trench border, the droplet left away. The black dashed lines in a) and c) represent the trench boundaries. The scale bars are 150 nm in (a), (b) and 500 nm in (c).

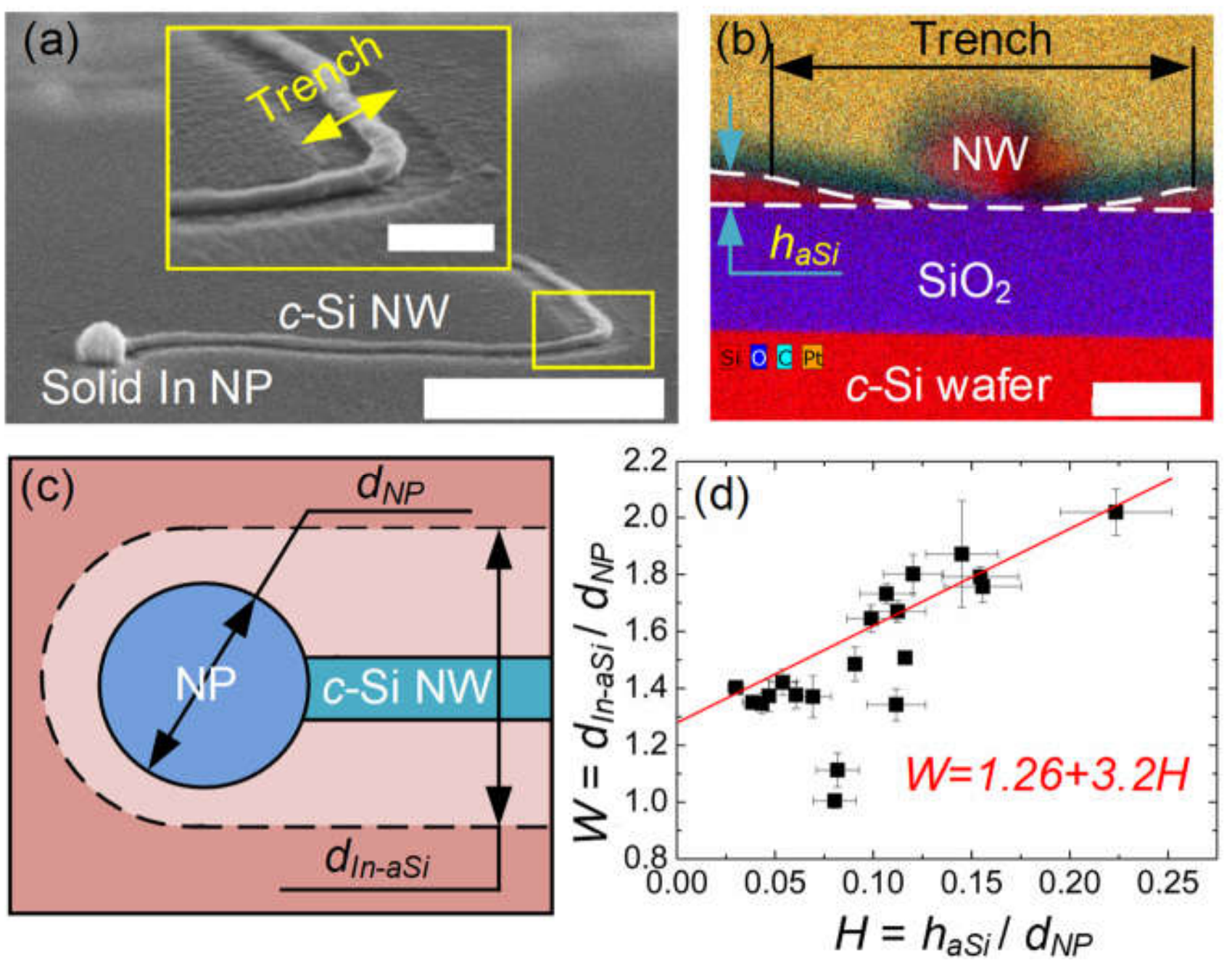

FIG. 3. Geometric investigation of the L-S systems. (a) SEM image of an IPSLS *c*-Si NW grown on $SiO_2$. The inset image provides a closer view of NW located in the center of the trench. (b) HAADF-STEM EDS relative map of the cross-sectional profile of a *c*-Si NW at the center of the trench obtained. $h_{aSi}$ denotes the *a*-Si:H thickness. (c) Top-view scheme of an IPSLS *c*-Si NW: $d_{NP}$ denotes the solid NP size, $d_{In\text{-}aSi}$ is the trench width. (d) Geometric analysis shows $W=d_{In\text{-}aSi}/d_{NP}$ has a linear relationship with $H=h_{aSi}/d_{NP}$. The scale bars in (a), inset of (a) and (b) are 1 μm, 200 nm and 80 nm, respectively.

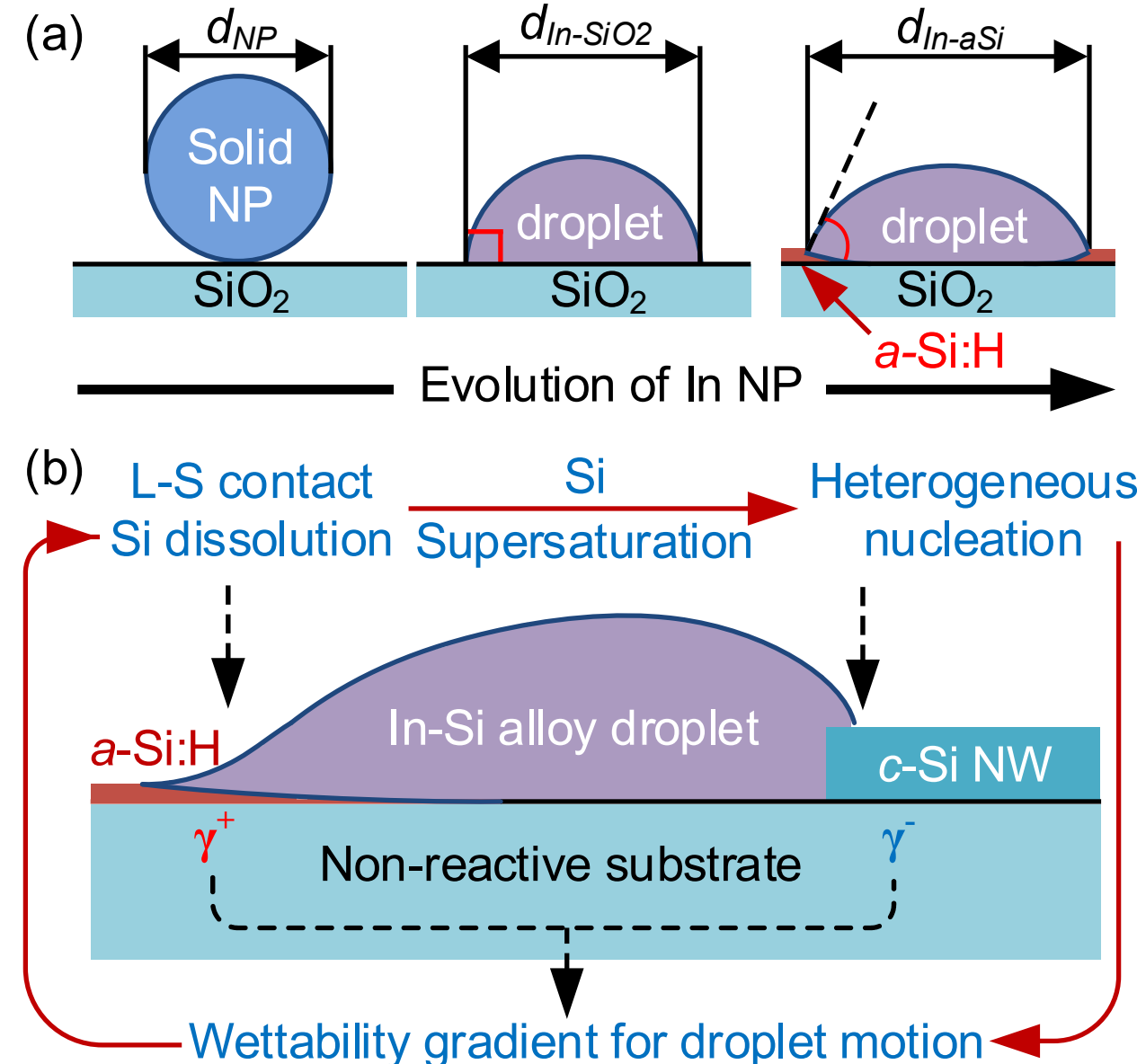

FIG. 4. (a) Evolution of an In NP from solid to liquid phase, from non-reactive spreading on bare $SiO_2$ to reactive spreading on *a*-Si:H coated $SiO_2$. (b) From side view, (i) In droplet reactively spreads and dissolves *a*-Si:H (high $\gamma^+$ surface), followed by the formation of In-Si alloy and Si supersaturation and heterogeneous nucleation, which is driven by the Gibbs free energy difference between *a*-Si:H and *c*-Si phases (denoted as $G_{a\mathrm{Si}}$ and $G_{c\mathrm{Si}}$, respectively); (ii) the droplet dewets the substrate surface at its rear side (low $\gamma^+$ surface), thereby a wettability gradient is built up between the front and rear edges of the droplet; (iii) the wettability gradient activates the In droplet flowing towards the *a*-Si:H coated area, which in consequence maintains the contact between the droplet and *a*-Si:H and enables a continuous NW growth.

**Supplementary Materials**

## 1. Experiments

*In-plane solid-liquid-solid Si NWs growth and geometric studies*: Indium was thermally evaporated on *c*-Si wafers having a 100 nm thermal oxide layer. Then In pads in size of 500 nm x 500 nm were defined by electron beam lithography. The nominal thickness of the pads was 5 nm or 50 nm. Then the sample was transferred to a 13.56 MHz RF-PECVD reactor, and treated by a standard process for the nanowire growth: (a) $H_2$ plasma treatment (100 sccm of $H_2$, 400 mTorr, RF-power density of 62 mW $cm^{-2}$ at substrate temperature of 300 °C for 5 minutes) to reduce the indium surface oxide; (b) *a*-Si:H deposition (5 sccm of pure $SiH_4$, 120 mTorr, RF power density of 24 mW $cm^{-2}$ at substrate temperature of 150 °C) in three different thicknesses of 9, 19 nm (on 5 nm In pads) and 30 nm (on 50 nm In pads). The *a*-Si:H thickness was calibrated by UV-vis spectroscopic ellipsometry (Jobin Yvon-MWR). (c) Annealing the sample at 450 °C in $H_2$ atmosphere (200 sccm, 1.9 Torr) for 10 minutes. We selected several successfully grown in-plane silicon nanowires from the samples for SEM (Hitachi S4800) observation.

The variation of In NP diameters was caused by the coalescence of NPs at the initial stage of nanowire growth. Fig. S1(a) shows an IPSLS Si NW grown from a 5 nm thick In pad. As seen in Fig. S1(b), the as-evaporated In pad was a discontinuous film with In NPs smaller than 80 nm. As a matter of fact, not all the In NPs were able to dissolve the *a*-Si:H coating layer and thereafter produce Si NWs, most of them were covered by the *a*-Si:H shells. As seen in Fig. S6, we carried out a simple test of the shells, which can be etched by $SF_6$ plasma rather than HCL. This provides a direct evidence that the shells were made of *a*-Si:H and are not due to the insufficient reduction of In surface oxide. Once a droplet broke its shell and initiated the NW growth (see the starting point), it could "rescue" and accumulate other In NPs in its movement path by breaking their shells. Thereafter, the final droplet got a much larger size (~160 nm) and was released from the In pad for NW growth (as seen in Fig. S1(c)). Thanks to this irregular coalescence, we managed to obtain different sizes of In droplets migrating on the *a*-Si:H for the Si NWs growth. In addition, as 32 nm *a*-Si:H was too thick for NPs of 5 nm In pads to produce silicon nanowires, we chose 50 nm In pads instead.

*In situ TEM experiment*: we use Aduro Thermal E-chips (Protochips Incorporated) for the *in situ* TEM observation of in-plane solid-liquid-solid growth. They consist in a ~50 nm thick amorphous $SiN_x$ membrane laying on a heating silicon carbide (SiC) support that contains 7x7 holes in diameter of 7 μm. The growth is observed over the holes where there is only the $SiN_x$ membrane. The temperature can be tuned continuously from room temperature to 1200 °C, based on Joule heating effect of the (doped) SiC by electrical current. An In thin film of ~200 nm (in shape of discontinuous islands) was deposited on the E-chip by thermal evaporation (see Fig. S2(a)). In order to decrease the density of In islands, the E-chip was dipped in 3% HCL solution for 3 minutes. After a ~1 min rinse in deionized (DI) water, the E-chip was placed on a hot plate at 125 °C for 2 minutes to evaporate the DI water on it, as shown in Fig. S2(b). Then the E-chip was transferred to the RF-PECVD reactor, and treated by a standard $H_2$ plasma exposure for 5 minutes (see the section In-plane solid-liquid-solid Si NWs growth for geometric analysis). Fig. S2(c) shows an SEM image of a membrane after $H_2$ plasma treatment. Note that the shape of the In particles evolved from irregular islands to spherical particles. After this SEM observation, the chip was returned to the PECVD system, and treated by: (a) a $H_2$ plasma exposure for 5 minutes; (b) 16 nm *a*-Si:H coating. Then, the E-chip was quickly transferred to TEM equipment (JEOL-2010F field emission electron microscope). The silicon nanowire growth was activated via the following process: small In droplets (up to several hundred nm) leaked from large In droplets (up to several μm), migrated on *a*-Si:H coated $SiN_x$ membrane and produced silicon nanowires, as shown in Fig. S3.

*STEM-HAADF EDX analysis*: chemical analyses have been performed in STEM-HAADF (scanning transmission electron microscope high-angle annular dark field) imaging mode of a 200kV Titan-Themis TEM/STEM electron microscope equipped with a Cs probe corrector and a ChemiSTEM Super-X detector. Prior to the analysis, the FIB technique was for the preparation of a cross-section lamella containing a Si NW inside the trench. The STEM-HAADF EDX chemical mapping was performed by considering the following elements of interest: the silicon Kα-1.73 keV ionization present also in the substrate and in the thermal oxide layer, the oxygen Kα-0.523 keV ionization edge coming also from the thermal oxide layer and the carbon Kα-0.277 keV ionization edge from the protective layer.

## 2. Temperature dependent In droplet surface tension

According to Ref. 41, the surface tension of an In droplet as a function of temperature T can be expressed as

$$\gamma_{In} = (568.0 - 0.04T - 7.08 \times 10^{-5}T^2 \pm 5) \times 10^{-5}\ N/cm. \quad (S1)$$

Thus, the In droplet surface tension at 350 °C and 400 °C equals 545±5 N/cm and ~ 541±5 N/cm, respectively.

## 3. Temperature dependent Si supersaturation for the heterogeneous nucleation from In droplet

According to Ref. 42, the Si chemical potential difference $\Delta\mu$ between In droplet in the supersaturated and the equilibrium states is expressed as

$$\Delta\mu = k_B T ln\left(\frac{C_{s,Si(In)}}{C_{0,Si(In)}}\right), \tag{S2}$$

where $k_B$ denotes the Boltzmann constant, $T$ the Kelvin temperature, $C_{s,Si(In)}$ and $C_{0,\mathrm{Si}(In)}$ the Si concentration in the supersaturated and saturated (i.e. in equilibrium state) In droplet, respectively, and $C_{s,Si(In)}/C_{0,\mathrm{Si}(In)}$ represents the supersaturation degree.

The heterogeneous nucleation energy barrier $\Delta G^*_{hetero}$ can be expressed as

$$\Delta G^*_{hetero} = \Delta G^*_{homo} f(\alpha) = [\chi \frac{(\Omega_{Si}\gamma_{Si-In})^3}{(\Delta\mu)^2}] f(\alpha), \tag{S3}$$

where $\Delta G^*_{homo}$ denotes the corresponding homogeneous nucleation energy barrier, $\chi$ denotes the shape factor of a nucleus (which equals $16\pi/3$ for a sphere and 32 for a cubic), $\Omega_{Si}$ denotes the volume of a Si atom, $\gamma_{Si\text{-}In}$ denotes the interfacial energy between the Si nucleus and the liquid In, $f(\alpha)$ denotes the contact angle ($\alpha$) factor between the Si nucleus and the substrate. In particular, the contact angle factor $f(\alpha)$ increases monotonically with the increasing $\alpha$.

Since the Si chemical potential difference should be higher than the nucleation barrier for the Si nucleation, by integrating Equations (S2) and (S3), we have Equation (2) in the manuscript

$$\begin{cases} \Delta\mu > \chi \dfrac{(\Omega_{Si}\gamma_{Si-In})^3}{(\Delta\mu)^2} f(\alpha) \\ \rightarrow \Delta\mu > \sqrt[3]{\chi f(\alpha)}\,\Omega_{Si}\gamma_{Si-In} \\ \rightarrow C_{s,Si(In)} > C_{0,Si(In)} \exp\,[\sqrt[3]{\chi f(\alpha)}\,\Omega_{Si}\gamma_{Si-In}/(k_B T)] \end{cases}$$

## 4. Zigzag growth of Sn catalyzed IPSLS Si or Ge NWs

According to Equation (2), due to the higher equilibrium concentration (i.e. solubility) of Si (or Ge) in Sn compared with those in In (i.e. $C_{0,Si(Sn)} > C_{0,Si(In)}$, $C_{0,Ge(Sn)} > C_{0,Si(In)}$) [1, 2], the Sn droplet has to intensify its spreading to accommodate more Si (or Ge) atoms to approach a higher critical supersaturation $C_{s,Si(Sn)}$ (or $C_{s,Ge(Sn)}$) for the nucleation. This is very similar to the growth of In catalyzed IPSLS Si NW at 350 °C, which favors the formation of the multi-faceted droplet-NW interfaces that renders higher probability of NW self-turning and risks of droplet mass loss along its trail.

## 5. Effect of the *a*-Si:H thickness on the droplet spreading

At a given growth temperature, considering ($d_{NP}$, $h_{aSi}$) as the *input* parameters to a SLS system, $d_{In\text{-}aSi}$ can be viewed as the *output*. Note that the NW diameter is also an *output* parameter, however since it is the trench but the NW that records the L-S contact during droplet flowing, this relationship can be given by

$$d_{In-aSi} = f(h_{aSi}, d_{NP}). \qquad \text{(S4)}$$

By transforming Equation (S4) into

$$d_{In-aSi}/d_{NP} = g(h_{aSi}/d_{NP}\,, 1). \qquad \text{(S5)}$$

Let $W=d_{In\text{-}aSi}/d_{NP}$, $H=h_{aSi}/d_{NP}$, Equation (S5) turns to be

$$W = g(H, 1), \qquad \text{(S6)}$$

which is used for the plotting and linear fitting in Fig. 3(d) and Equation (3).

For a droplet with a constant volume under constant pressure and temperature, the thicker the *a*-Si:H coating layer is, the more Si atoms the droplet has to dissolve in order to deplete the *a*-Si:H beneath it for the *c*-Si precipitation (see Fig. 3(b)). The Gibbs free energy change of the droplet (more precisely the liquid In-Si alloy) $\Delta G_{drop}$ is expressed as [3]

$$\Delta G_{drop} = \mu_{In}\Delta n_{In} + \mu_{Si}\Delta n_{Si}, \qquad \text{(S7)}$$

where $\mu_{In,Si}$ denotes the chemical potential of In and Si, respectively; $\Delta n_{In,Si}$ denotes the molar mass change of In and Si, respectively. Thereby, once the *a*-Si:H gets thicker, the droplet will have an excess Gibbs free energy contributed by the increment of $\Delta n_{si}$, which will render the droplet flattening to increase its surface free energy for approaching to a new equilibrium state, that is, the droplet spreading will be enhanced.

## 6. Supporting Figures

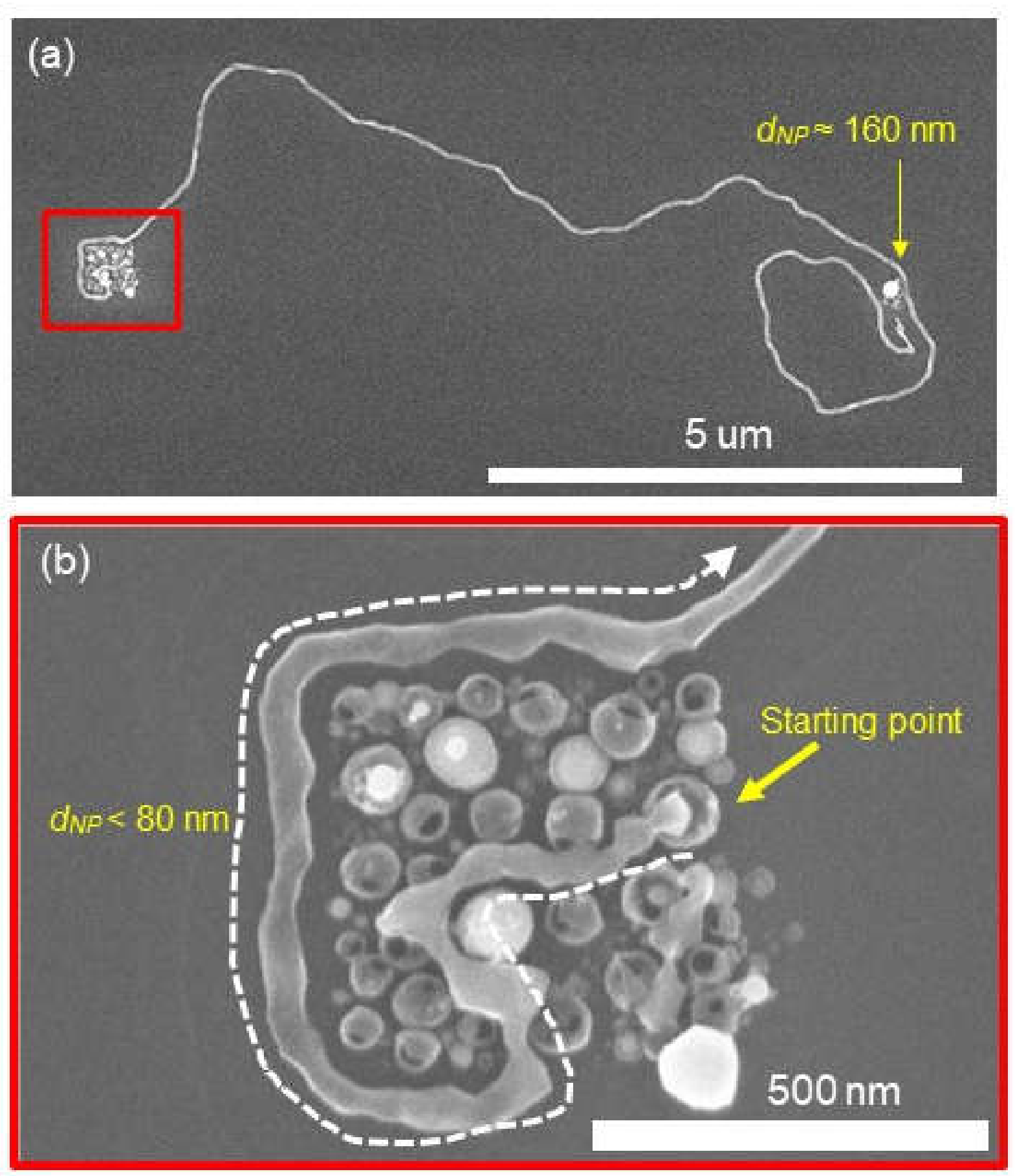

FIG. S1. An example of In NPs coalescence at the initial stage of an in-plane solid-liquid-solid Si NW growth. (a) SEM image of an in-plane Si NW grown from a 5 nm thick In pad. (b) Zoom-in area in (a), part of the evaporated In NPs were not activated to dissolve the *a*-Si:H coating layer. However, one In droplet broke the shell and initiated the NW growth (see the starting point). Interestingly, this moving droplet managed to rescue and accumulate In from other NPs along its path by breaking their shells, thereafter the final droplet (whose size is much larger than those escaped from the broken shells) was released for NW growth. Thanks to this random coalescence, we managed to obtained different sizes of In droplets migrating on the *a*-Si:H for the Si NWs growth.

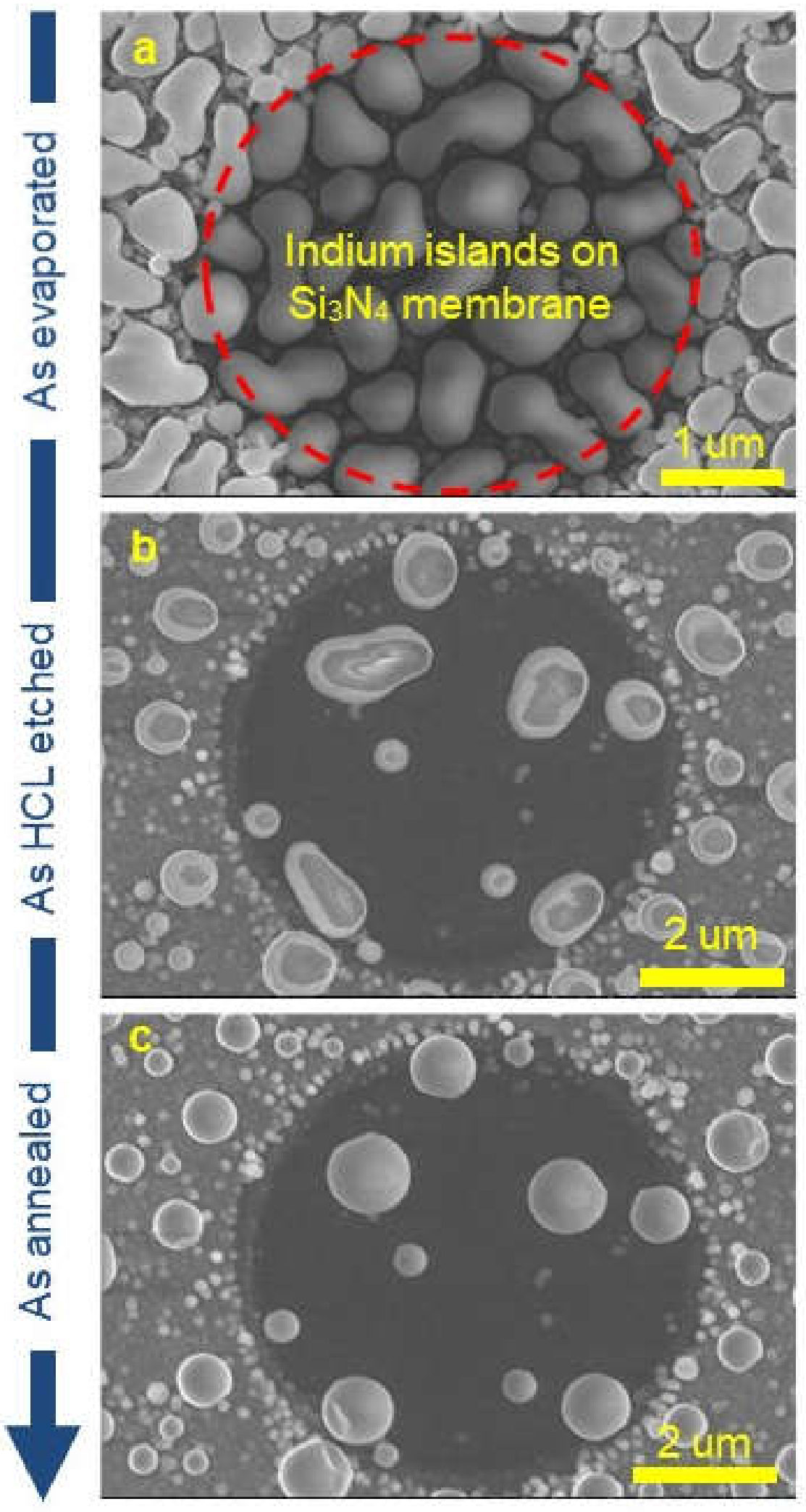

FIG. S2. In NPs preparation on a TEM heating membrane. (a) SEM image of a membrane after 200 nm In thin film evaporation, the $SiN_x$ hole is in the dark center part. (b) SEM image of a membrane after dipping in 3% HCl solution for 3 minutes. (c) SEM image of a membrane after standard $H_2$ plasma treatment for 5 minutes.

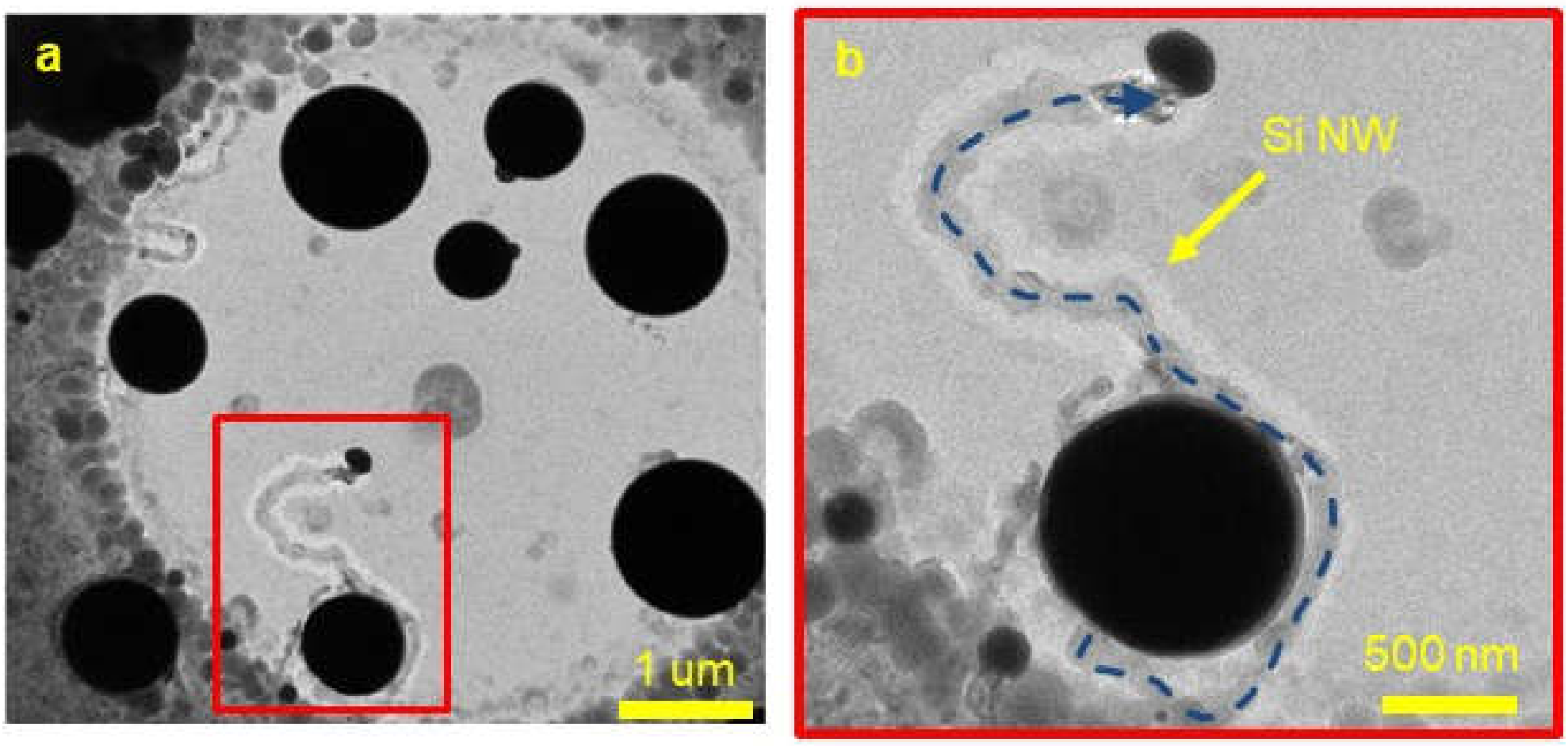

FIG. S3. (a) A small In droplet leaked from a large stationary In NP, migrated on *a*-Si:H coated $Si_3N_4$ membrane and produced a Si NW. (b) a zoom-in view of the Si NW in (a).

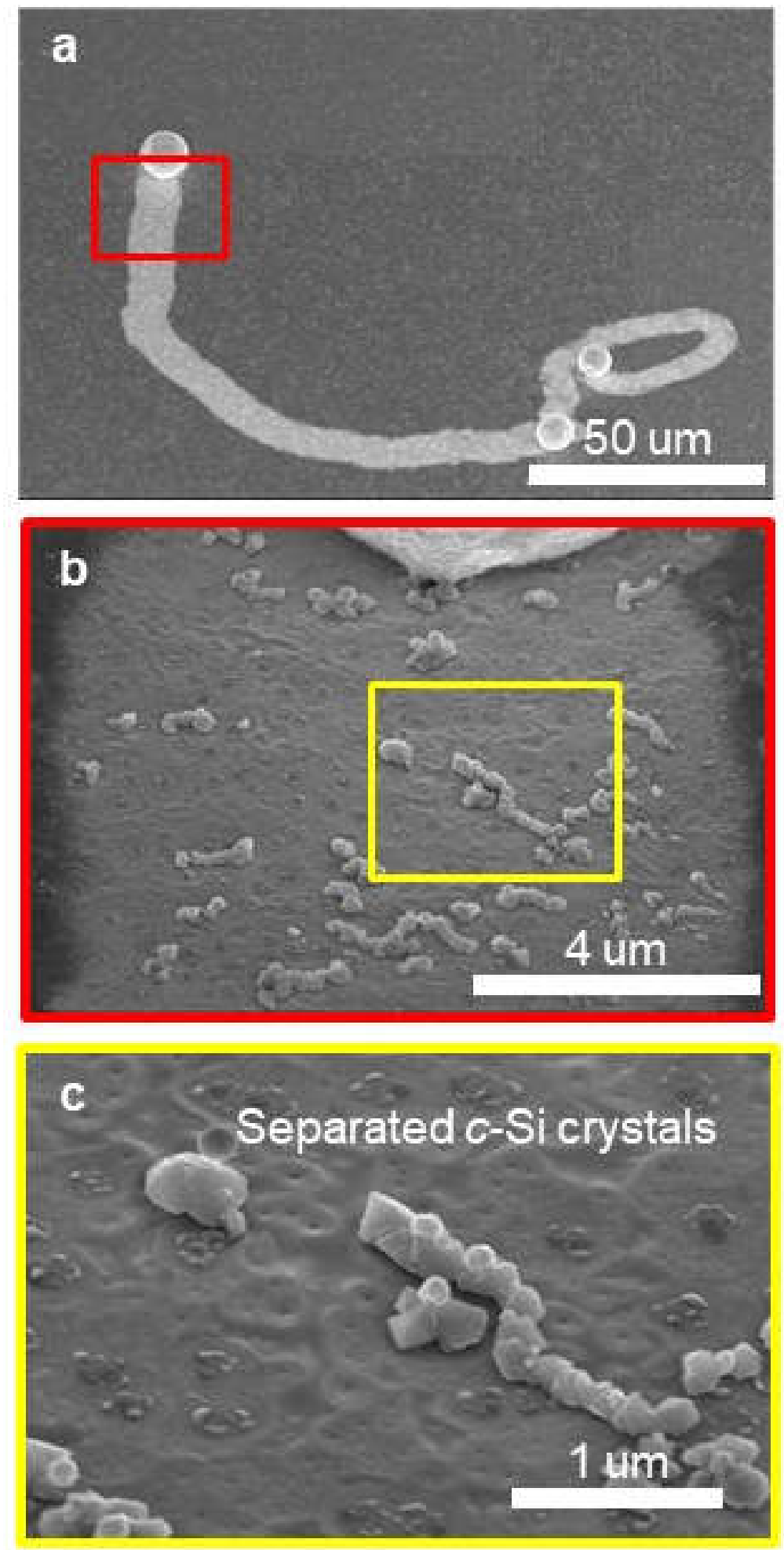

FIG. S4. (a) Migration of a 10-μm In droplet on *a*-Si:H coated substrate, leaving separated silicon crystals on its trace. This indicates that the wettability gradient due to a progressive substrate surface change (transition from *a*-Si:H on a raw substrate to *c*-Si on a raw substrate) is the requirement for In droplet motion, while *c*-Si NW growth (i.e. continuous one-dimensional crystal growth) is a particular case. (b) and (c) the zoom-in view of the separated *c*-Si crystals.

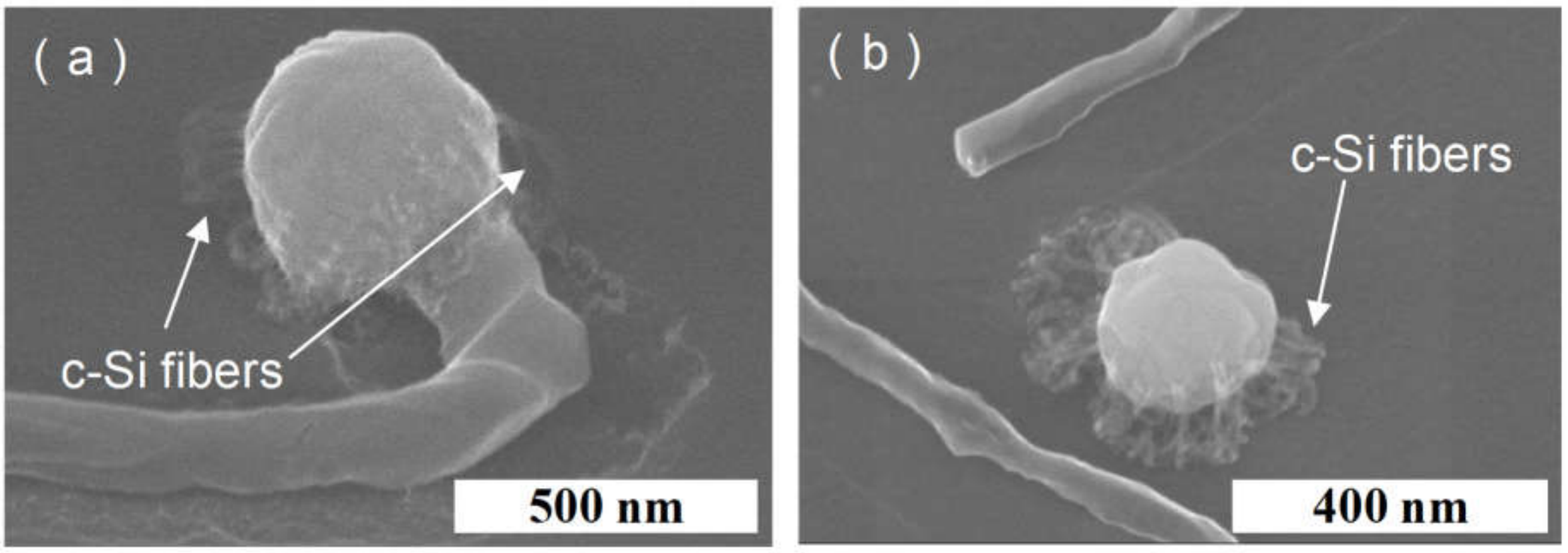

FIG. S5. *c*-Si fibers precipitated from the droplet during the solidification of the In-Si alloy, of which the “arms” observed in the TEM video 1 is probably made (see Fig. 1(b) $t = 4$ s in the manuscript).

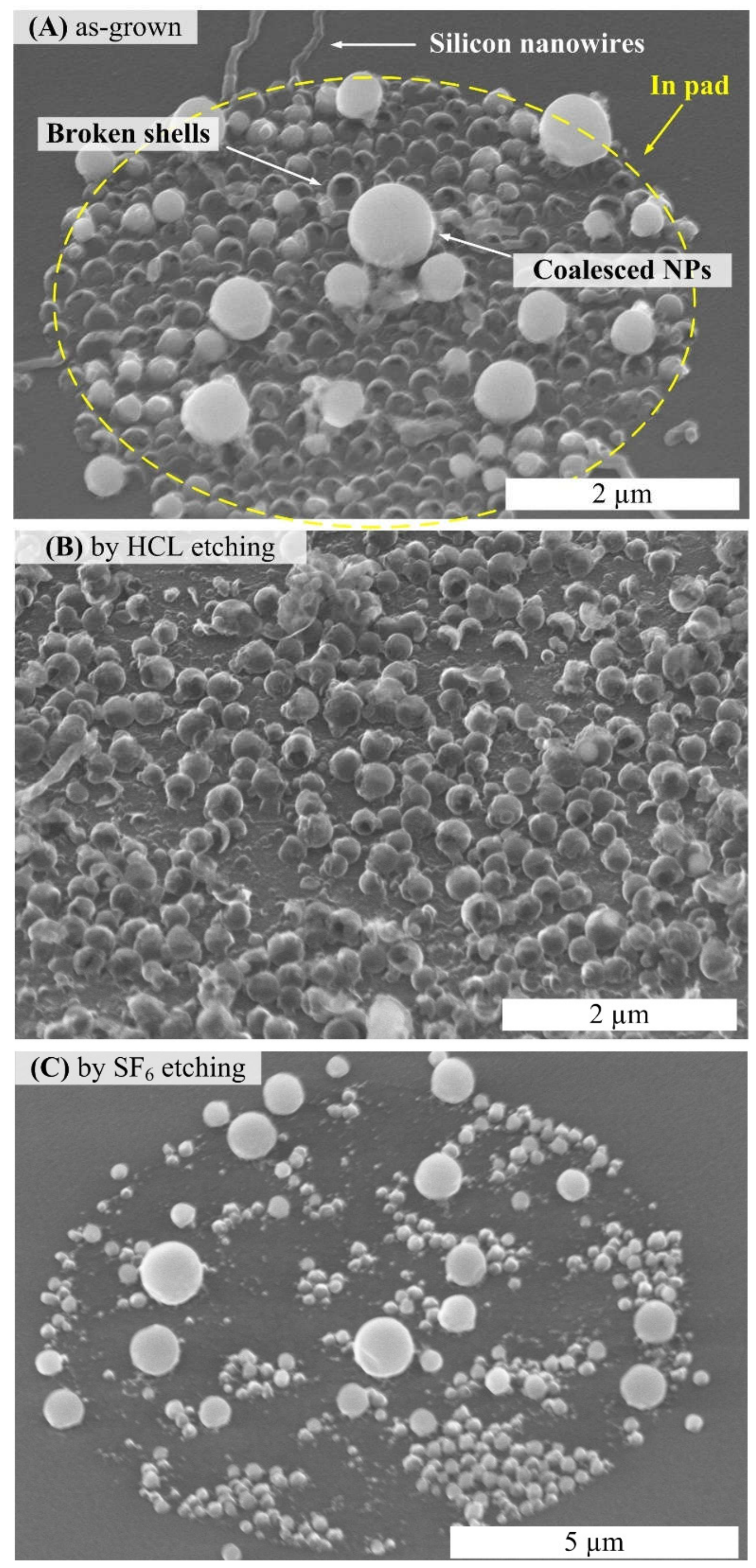

FIG. S6. Characterization of the shells formed at the initial stage of the IPSLS Si NW growth. (a) The shells formed after the growth of IPSLS Si NW out of the In pad. (b) The shells treated by HCL etching. (c) The shells treated by $SF_6$ plasma etching.

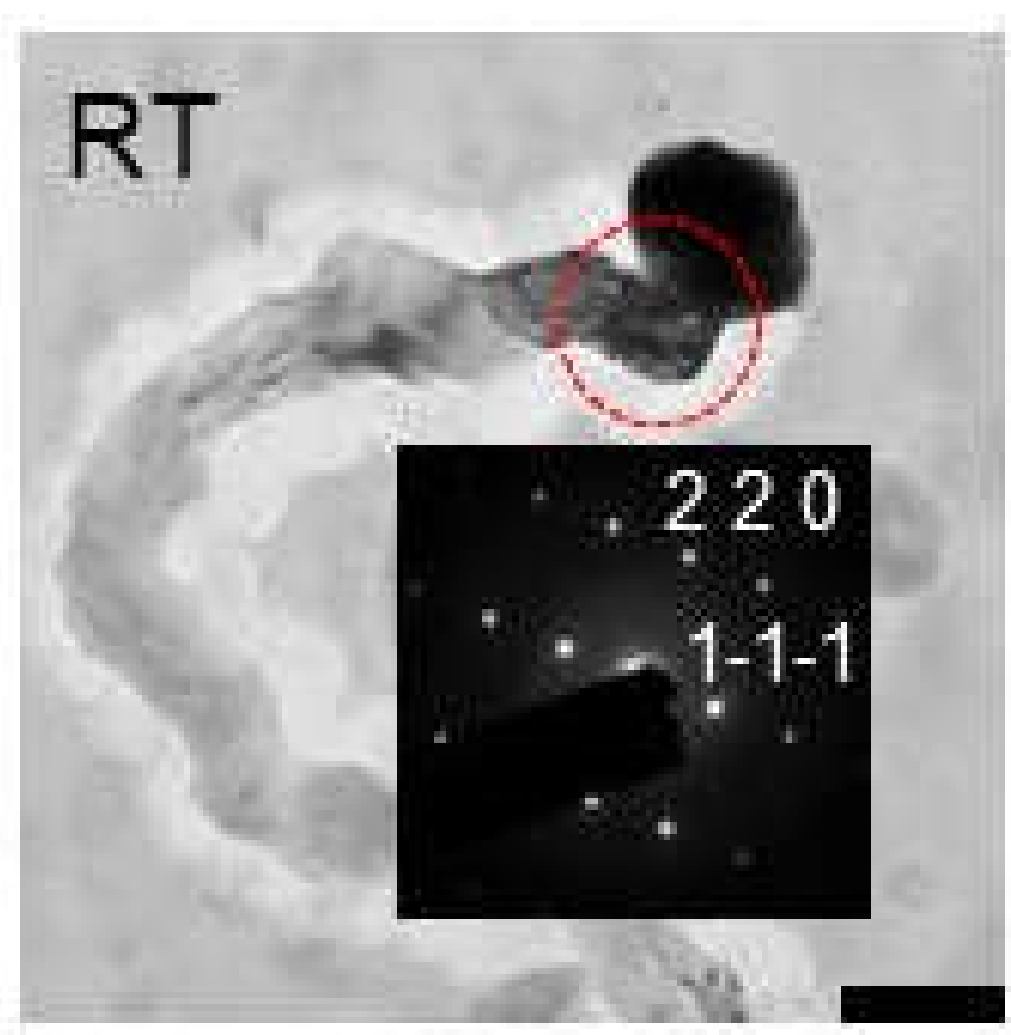

FIG. S7. TEM image taken at room temperature of a Si NW grown on a $SiN_x$ membrane at 400 °C. Inset: selected area electron diffraction pattern in [1-12] zone showing the local orientation of *c*-Si.

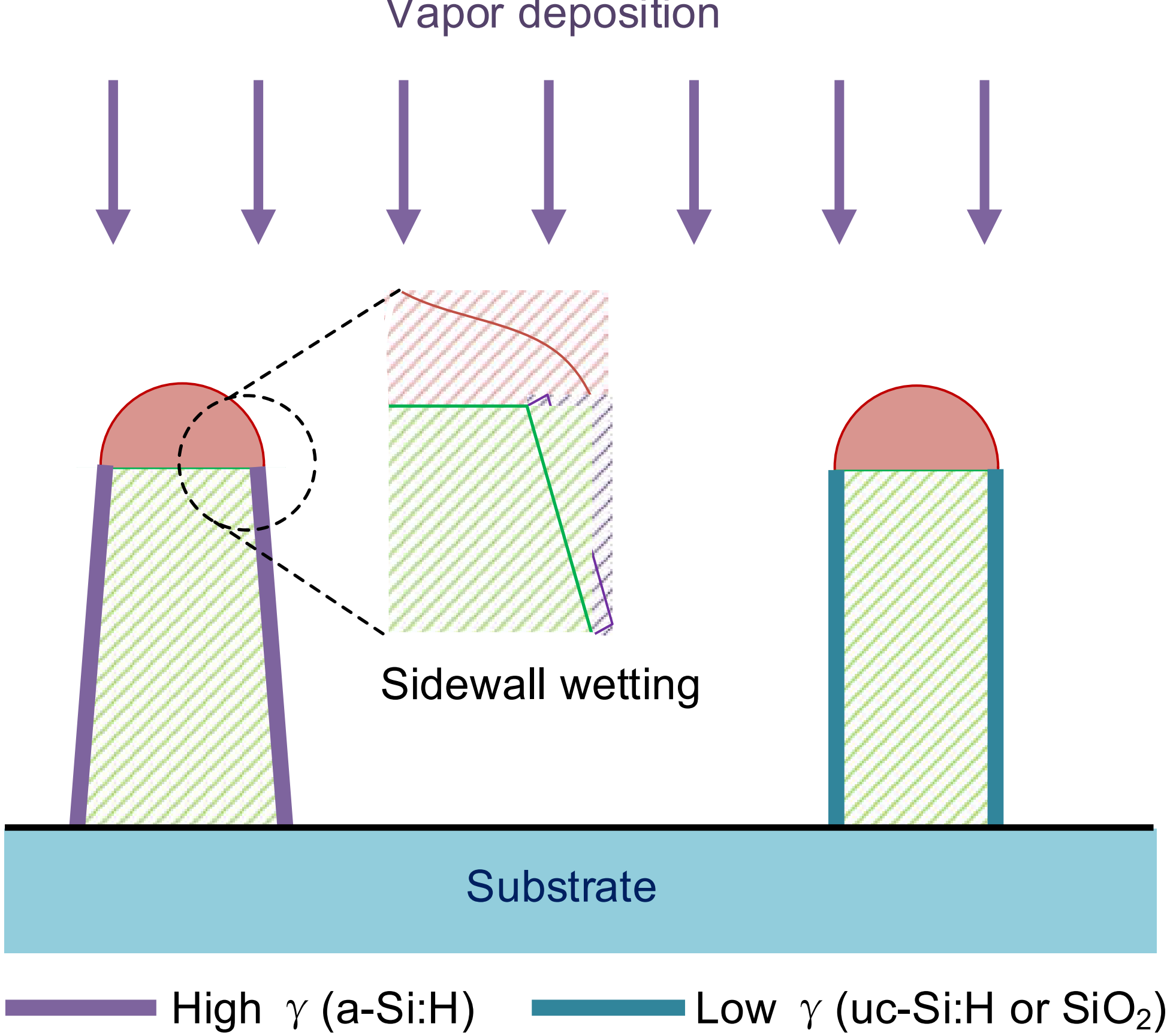

FIG. S8. NW tapering during the catalytic droplet sidewall wetting: (left) schematic of the mechanism of the Si nanowire surface wetting by the catalytic droplets; (right) the possible methods to suppress this wetting behavior by oxidizing the NW sidewall (see Ref. 46) and depositing *μc*-Si:H instead of *a*-Si:H for the VLS growth (unpublished results, *μc*-Si:H does not react with metal catalyst like In or Sn, see Fig. S11). In addition, it is reported that the PECVD deposition of *a*-Si:H with diluted silane at higher temperature helps suppressing the droplet sidewall wetting and enable long Si NWs growth. [4] The optimized recipe corresponds to a low deposition rate [5], which is regarded as an alternative way to suppress the droplet sidewall wetting. This result is consistent with the conclusion that thicker *a*-Si:H renders stronger droplet reactive wetting.

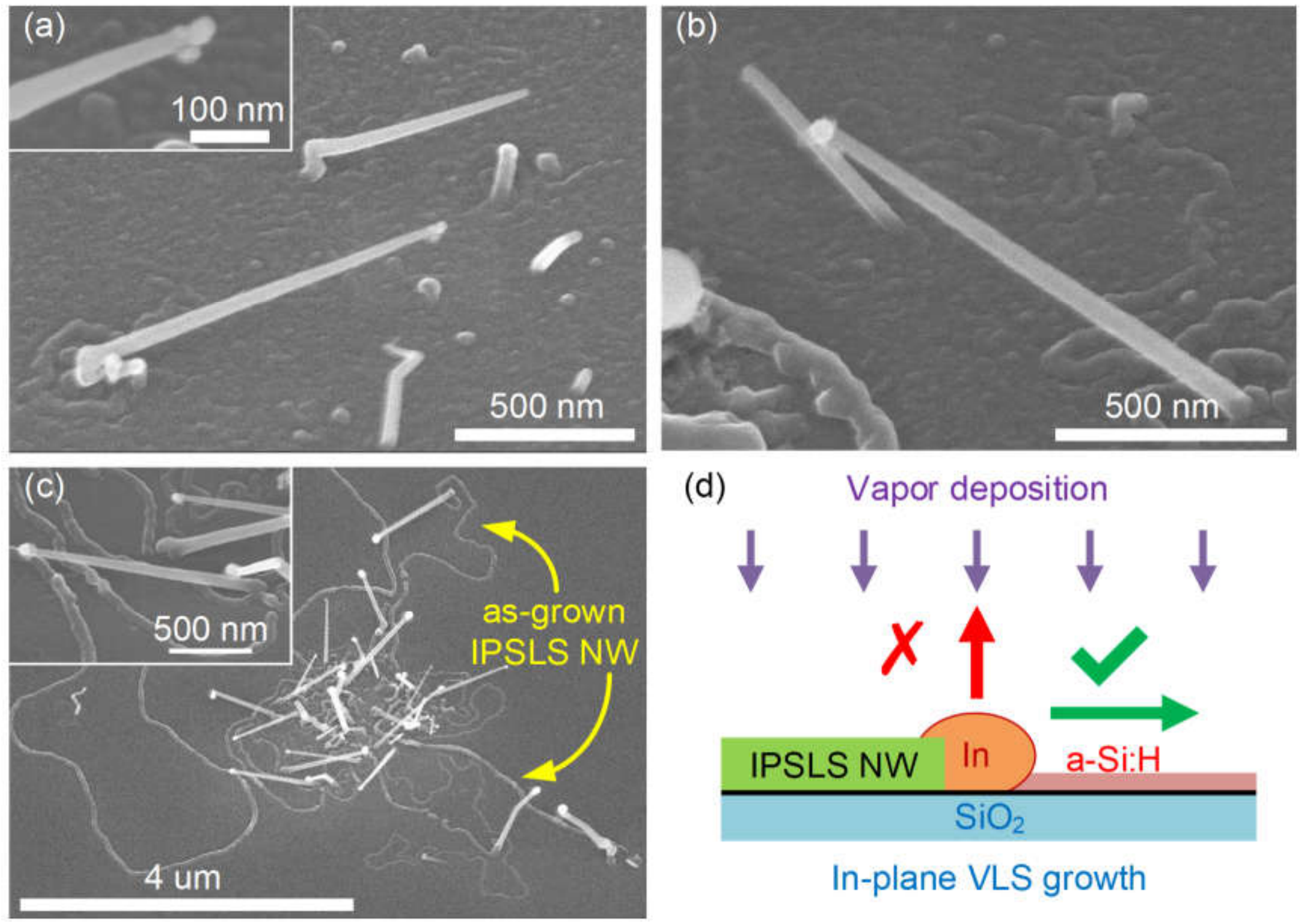

FIG. S9. In catalyzed VLS growth of lateral Si nanowires on *a*-Si:H coated substrate. (a-c) SEM images of VLS grown lateral Si nanowire on the substrates after the IPSLS process, (d) the schematic of the in-plane growth mechanism: the *a*-Si:H coating layer enhances substrate surface wettability, thereby prohibits the detachment of the In droplets from the substrate and gives rise to the in-plane VLS growth, which is in accordance with the theoretical simulations by Schwarz et al. (see Refs. 48, 49). The VLS growth condition of In catalyzed Si NWs is: 150 sccm $H_2$, 3.3 sccm $SiH_4$, total pressure of 1 Torr, RF power density of 38 mW/cm$^2$, 30 mins.

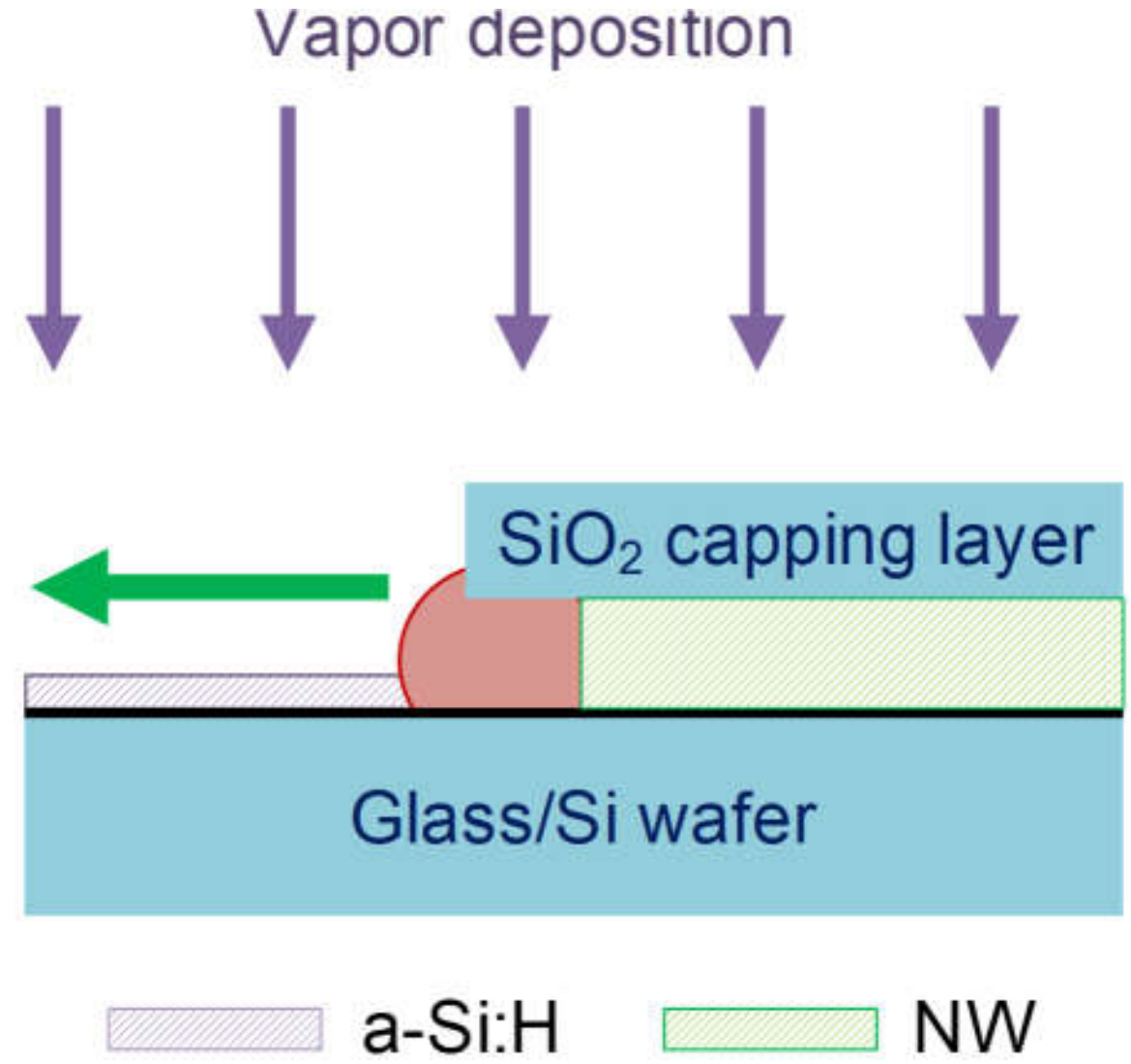

FIG. S10. Schematic of the mechanism of the guided growth by nanochannels (see Ref. 18): before the droplet moves out of the nanochannel, there has been a thin layer of *a*-Si:H deposited on the substrate [6], thus the NW will grow laterally in the same manner as discussed in Fig. S9.

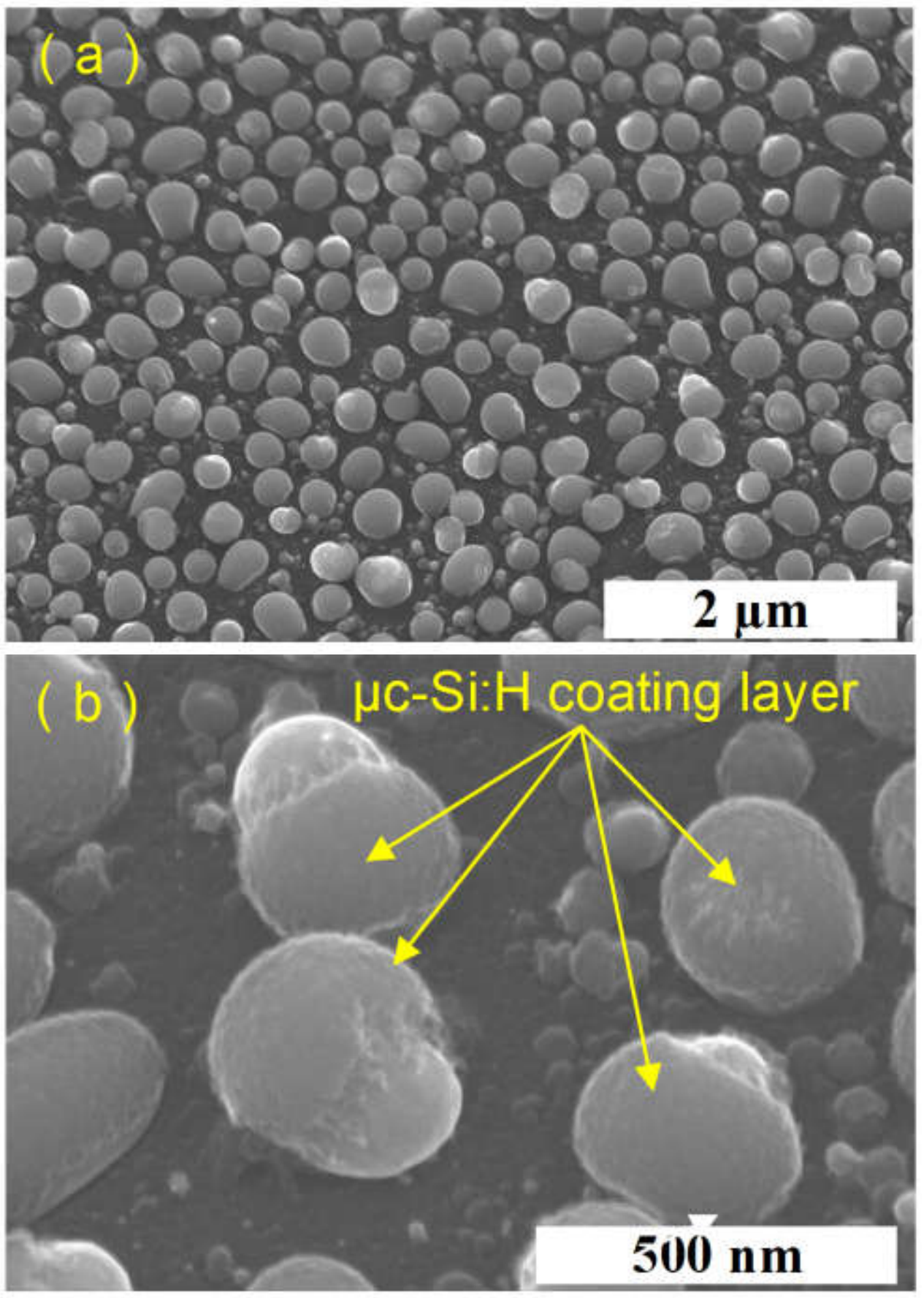

FIG. S11. An IPSLS growth process by depositing *μc*-Si:H instead of *a*-Si:H, with no NW grown, which demonstrates that In is not able to dissolve Si atoms from *μc*-Si:H. The deposition condition of *μc*-Si:H is: 200 sccm $H_2$, 0.3 sccm $SiH_4$, total pressure of 1.9 Torr, RF power density of 255 mW/cm$^2$, 20 mins.

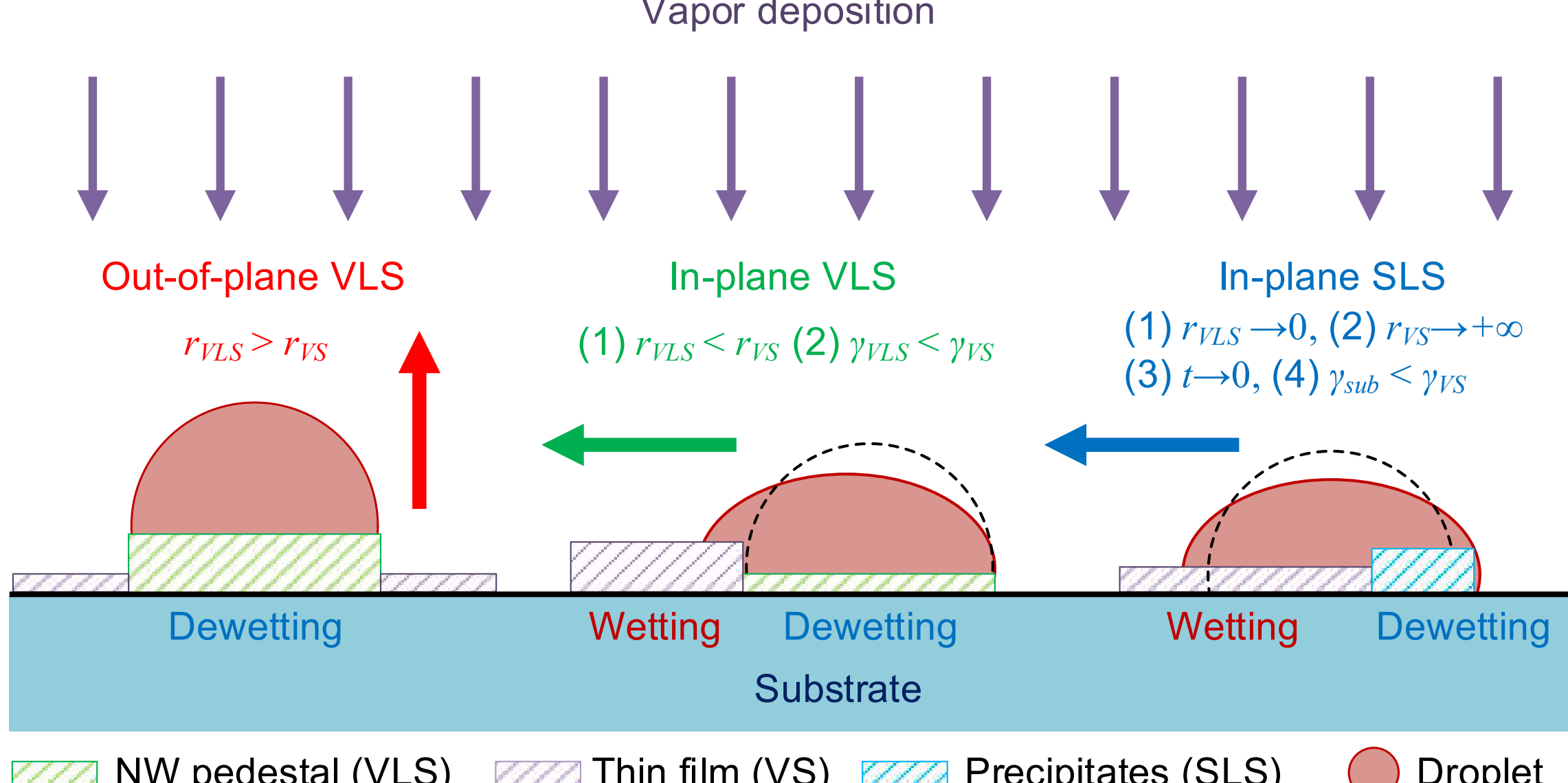

FIG. S12. Evolution of catalytic droplet motion under vapor deposition for NWs growth. (a) If the formation of the NW pedestal is faster than the thin film deposition on the substrate (i.e. $r_{VLS} > r_{VS}$), the droplet will be detached from the substrate surface and drive an out-of-plane NW growth. (see Ref. 49) (b) If the thin film grows faster than the nanowire (i.e. $r_{VLS} < r_{VS}$), meanwhile with a higher surface energy than the NW pedestal (i.e. $\gamma_{VLS} < \gamma_{VS}$), the droplet will wet the thin film, which results in the VLS in-plane NW growth. An example is the Pd mediated in-plane VLS Si NWs, while the In droplets drive the out-of-plane ones under the same condition.[7] This is probably because the Si nucleation rate in Pd is lower than that in In, as the solubility of Si in Pd is larger than that in In.[8] Being a variant, the condition of $r_{VLS} < r_{VS}$ *can also be realized by pre-coating a thin film before the out-of-plane growth (like those in Figs. S9 and S10).* (c) If a thin film with a higher wettability compared with the substrate is deposited in an infinitively high rate and in an infinitively short period during which the NW pedestal is hardly grown (i.e. $r_{VLS} \rightarrow 0$, $r_{VS} \rightarrow +\infty$, $t \rightarrow 0$, $\gamma_{VS} > \gamma_{SUB}$, $\gamma_{VS} > \gamma_{SLS}$), which can be realized technically by depositing the thin film deposition prior to the NW growth, an IPSLS growth will take place.

## 7. Supporting Videos

Video SV1: an *in situ* transmission electron microscopy video of IPSLS Si nanowire growth at 350°C (1 frame $s^{-1}$).

Video SV2: an *in situ* transmission electron microscopy video of IPSLS Si nanowire growth at 400°C (1 frame $s^{-1}$).